\documentclass[floats,floatfix,amssymb,prd,twocolumn,superscriptaddress,nofootinbib]{revtex4-1}
	
\usepackage{subcaption}
\usepackage{ragged2e}
\DeclareCaptionJustification{justified}{\justifying}
\makeatletter
\newcommand{\subsetsim}{\mathrel{\mathpalette\subset@sim\relax}}
\newcommand{\subset@sim}[2]{%
  \vtop{\offinterlineskip\m@th
    \ialign{\hfil##\cr
      $#1\subset$\cr\noalign{\kern0.5pt}\scalebox{0.9}{$#1\sim$}\cr
    }%
  }%
}
\makeatother

\usepackage{amssymb,amsmath,verbatim,mathtools,needspace,enumitem,etoolbox,graphicx,physics,microtype,afterpage,bm}
\usepackage[dvipsnames, usenames]{xcolor}
\definecolor{linkcolor}{rgb}{0.0,0.3,0.5}
\usepackage{booktabs}
\usepackage[unicode, colorlinks=true, linkcolor=linkcolor, citecolor=linkcolor, filecolor=linkcolor,urlcolor=linkcolor, pdfusetitle]{hyperref}
\usepackage[all]{hypcap}
\usepackage[T1]{fontenc}
\usepackage[utf8]{inputenc}
\usepackage{tabularx}
\usepackage{float}
\usepackage{empheq}

\newcommand{\La}{\mathcal{L}}
\newcommand{\unmezzo}{\frac{1}{2}}
\newcommand{\F}{{\rm F}}

\usepackage{multirow}
\usepackage{pifont}
\usepackage{lmodern}

\usepackage{multirow}

\allowdisplaybreaks
\usepackage{tikz}
\usepackage{color}
\usepackage{framed}
\usepackage{hyperref}
\hypersetup{colorlinks, citecolor=bluscuro, linkcolor=black, urlcolor=bluscuro}
\definecolor{rossos}{cmyk}{0,1,1,0.55}
\definecolor{bluscuro}{rgb}{0.15, 0.2, .85}
\definecolor{bluchiaro}{cmyk}{1,.3,0.,0.1}
\definecolor{ForestGreen}{rgb}{0.13, 0.55, 0.13}
\definecolor{azure}{rgb}{0.0, 0.5, 1.0}
 
\usepackage{slashed}

\def\nn{\nonumber}

\def\bea{\begin{eqnarray}}
\def\eea{\end{eqnarray}}

\def\d{{\mathrm{d}}}

\newcommand{\bs}{\begin{subequations}}
\newcommand{\es}{\end{subequations}}

\newcommand{\be}{\begin{equation}}
\newcommand{\ee}{\end{equation}}
\renewcommand{\d}{{\rm d}}

\def\lsim{\mathrel{\rlap{\lower4pt\hbox{\hskip0.5pt$\sim$}}
    \raise1pt\hbox{$<$}}}         
\def\gsim{\mathrel{\rlap{\lower4pt\hbox{\hskip0.5pt$\sim$}}
    \raise1pt\hbox{$>$}}}         

\renewcommand{\d}{{\rm d}}
\newcommand{\eff}{{\rm eff}}

\makeatletter
\def\l@subsubsection#1#2{}
\makeatother

\newcommand{\sapienza}{Dipartimento di Fisica, Sapienza Università 
	di Roma, Piazzale Aldo Moro 5, 00185, Roma, Italy}
\newcommand{\infn}{INFN, Sezione di Roma, Piazzale Aldo Moro 2, 00185, Roma, Italy}

\begin{document}
\title{
Compact objects in and beyond the Standard Model\\ from non-perturbative vacuum scalarization
}

\begin{abstract}
We consider a theory in which a real scalar field is Yukawa-coupled to a fermion and has a potential with two non-degenerate vacua. If the coupling is sufficiently strong, a collection of $N$ fermions deforms the true vacuum state, creating energetically-favored false-vacuum pockets in which fermions are trapped.
We embed this model within General Relativity and prove that it admits self-gravitating compact objects where the scalar field acquires a non-trivial profile due to non-perturbative effects.
We discuss some applications of this general mechanism: i)~\emph{neutron soliton stars} in low-energy effective QCD, which naturally happen to have masses around $2M_\odot$ and radii around $10\,{\rm km}$ even without neutron interactions; ii) \emph{Higgs false-vacuum pockets} in and beyond the Standard Model; iii)~\emph{dark soliton stars} in models with a dark sector. In the latter two examples, we find compelling solutions naturally describing centimeter-size compact objects with masses around $10^{-6}M_\odot$, intriguingly in a range compatible with the Optical Gravitational Lensing Experiment (OGLE) + Hyper Suprime-Cam (HSC) microlensing anomaly.
Besides these interesting examples, the mechanism of non-perturbative vacuum scalarization may play a role in various contexts in and beyond the Standard Model, providing a support mechanism for new compact objects that can form in the early universe, can collapse into primordial black holes through accretion past their maximum mass, and serve as dark matter candidates. 
\end{abstract}

\author{Loris Del Grosso}
\affiliation{\sapienza}
\affiliation{\infn}


\author{Paolo Pani}
\affiliation{\sapienza}
\affiliation{\infn}

\author{Alfredo Urbano}
\affiliation{\sapienza}
\affiliation{\infn}

\maketitle

{
  \hypersetup{linkcolor=black}
}

\section{Introduction}\label{sec:intro}

Non-topological solitons (NTSs) are classical solutions of non-linear field theories, stabilized by the existence of a conserved Noether charge~\cite{Lee:1991ax}. They are naturally studied in the context of extended and possibly macroscopic configurations. For this reason, several NTSs, such as Q-balls~\cite{Coleman:1985ki}, quark nuggets~\cite{Witten:1984rs, Bai:2018dxf}, and dark photon stars~\cite{Gorghetto:2022sue}, have been often associated to non-particle dark matter candidates (see e.g.~\cite{Gross:2021qgx, Ponton:2019hux, Ansari:2023cay, Hong:2020est}). 

\begin{figure}[t] 
\centering
\includegraphics[width=1\linewidth]{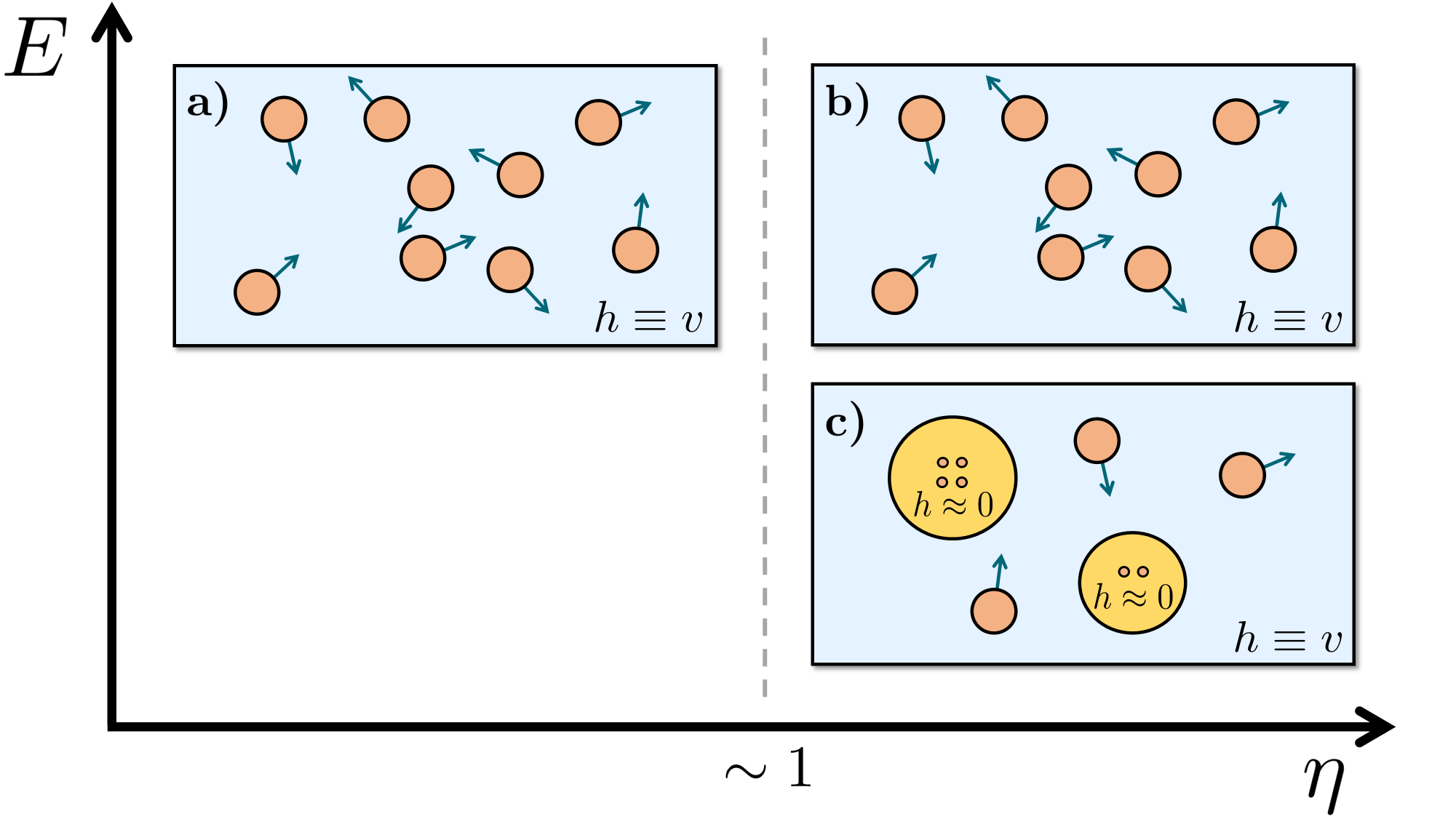}%
\caption{
   Sketch of the mechanism, showing the energy $E$ of different configurations as a function of $\eta$, a fundamental parameter defined in~Eq.~\eqref{dimensionless_parameters}. The bigger $\eta$ the more the theory becomes strongly coupled.
   The orange balls represent massive fermions, whereas the yellow balls represent the false vacuum pockets of the scalar field in which the fermions are massless (and therefore depicted by smaller orange balls).
   \textbf{(a)} Standard ground state of the system.
   \textbf{(b), (c)} Whenever $\eta \gtrsim 1$, it is energetically convenient for the system to trap a fraction of fermions in false vacuum pockets. For $\eta \lesssim 1$, equilibrium configurations describing bound false vacuum pockets do not exist.}
   \label{fig:illustration}
\end{figure}

In this work, we consider a simple model-theory for NTSs in which a real scalar field $h$ is coupled to a fermion through a Yukawa coupling.
The scalar potential features the typical Mexican-hat shape, with two minima in $h = \pm v$ and a maximum in $h = 0$. However, the presence of the fermionic condensate $S = \langle \bar{\psi}\psi \rangle$ effectively modifies the shape of the scalar potential (see Fig.~\ref{fig:potential}). In particular, if the Yukawa coupling $f$ is sufficiently strong (or the scalar quartic $\lambda$ sufficiently small), the point $h = 0$ becomes an actual (local) minimum of the theory. The latter condition is mathematically formulated by requiring that the parameter $\eta = f / (2\lambda)^{1/4} \gtrsim 1$. In this regime, we show the existence of a bound NTS where the scalar field interpolates between $h = 0$ (false vacuum) and $h = v$ (true vacuum)\footnote{Without loss of generality, we are assuming that the asymptotic (scalar) ground state of the Universe is $h = v$. One can also consider a solitonic solution that connects $h = 0$ with $h = -v$, but then we would have a topological soliton (because the asymptotic value of the scalar field in the solution would be different from its value in the rest of the Universe).}, working in a fully relativistic approach.

The physical implication of the latter result is that, if $\eta \gtrsim 1$, a collection of $N$ fermions is able to deform the $h \equiv v$ ground state, giving rise to NTSs that describe false vacuum pockets, in which the $N$ fermions are trapped (see Fig.~\ref{fig:illustration}). The latter are unable to escape (at least classically) because their energy is lower than $N m_f$, the rest energy of $N$ free fermions ($m_f$ is the fermion bare mass).
Thus, the energy of an everywhere uniform configuration $h \approx v$ becomes higher than the energy of a configuration that shows false vacuum pockets here and there. Since NTSs are intrinsically non-perturbative states (see Sec.~\ref{sec_thestim}), the ground state of the system acquires a nontrivial scalar profile by means of non-perturbative effects. We dub this mechanism \textit{non-perturbative vacuum scalarization}. 

Investigations in a similar spirit were recenly carried out by Ref.~\cite{Balkin:2023xtr} in the context of exotic neutron stars.

The primordial formation for NTSs has been investigated by many authors (see e.g.~\cite{Frieman:1988ut, Griest:1989bq, Cottingham:1991jy, Bai:2022kxq}), pointing out two main formation mechanisms, the \textit{solitosynthesis} and the \textit{solitogenesis}. In the former scenario, NTSs are formed by the fusion of $N$ free fermions. This implies a non-zero cosmic asymmetry since it is necessary to accumulate a net number $N$ of fermions over antifermions in a given region of space. In the latter scenario, a relic abundance of NTSs is produced through a first- or second-order cosmological phase transition. Recently, Ref.~\cite{Bai:2022kxq} reviewed these formation mechanisms in detail, showing that in certain cases NTSs can dominate the dark matter abundance.
A further formation channel is provided from the Yukawa interaction present in Eq.~\eqref{theory_fund}, as long as it is enough long-range. In this case, fermions undergo clustering and structure formation even in a radiation-dominated era~\cite{Domenech:2021uyx, Domenech:2023afs, Flores:2023zpf}.

Phenomenological imprints of NTSs could be found through gravitational waves~\cite{Lozanov:2023aez, Banks:2023eym} and lensing observations.
Besides being natural non-particle dark matter candidates, these solutions are also well-motivated examples of exotic compact objects~\cite{Giudice:2016zpa} and might share similar phenomenology with
ordinary compact objects such as black holes and neutron stars~\cite{Cardoso:2019rvt}.

\section{Model for relativistic NTSs}\label{sec:setup}
\begin{figure}[t] 
\includegraphics[width=1\linewidth]{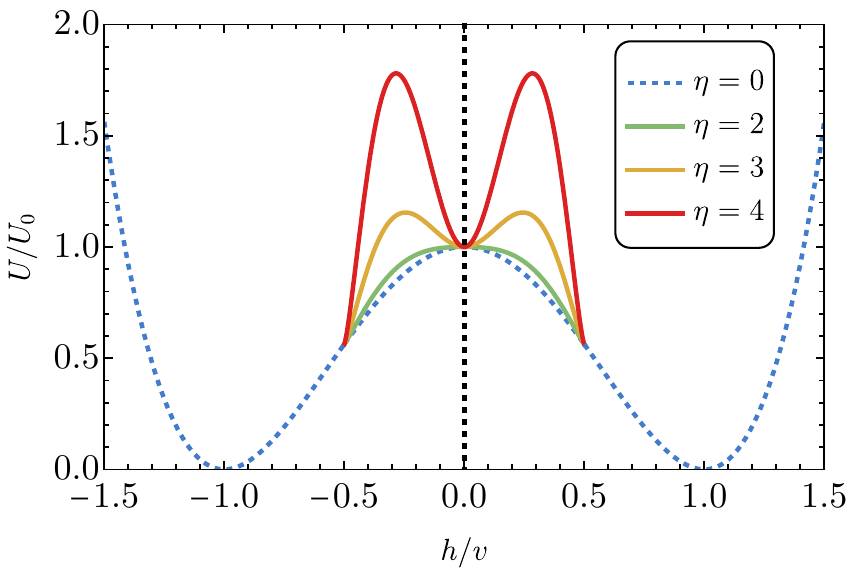}
	\caption{Effective scalar potential (see Eq.~\eqref{eq:effective_potential}) normalized with respect to $U_0 = \lambda v^4 / 16$ as a function of $h/v$ for various values of $\eta$ in the Newtonian regime. In this regime, the zeros of the Fermi momentum are $h/v = \pm \tilde{\omega}_\F$ (for $|h|/v > \tilde{\omega}_\F$ the fermions are removed from the theory).  We fixed a representative value $\tilde{\omega}_\F = 0.5$. We notice that when $\eta \gtrsim 1$ the effective potential develops a local minimum in $h = 0$. The shape of the potential is qualitatively the same also in the fully relativistic regime.}
	\label{fig:potential}
\end{figure}

Our starting point is the following theory, in which a (real) scalar boson $h$ and a Dirac fermion $\psi$ are minimally coupled to Einstein gravity\footnote{We use the signature $(-,+,+,+)$ for the metric and adopt natural units ($\hbar = c = 1$). With the normalization used for the fermionic kinetic term, the Dirac matrices have an extra $-i$ factor with respect to the usual ones but satisfy the usual relation $\{\gamma^\mu, \gamma^\mu\} = 2 g^{\mu\nu}$. 
The covariant derivative $D_\mu$ takes into account the spin connection
of the fermionic field. We are neglecting interactions with gauge fields, but it is straightforward to generalize our model including them.}, 
\begin{align}\label{theory_fund}
    \La &= \frac{R}{16\pi G} - \frac{1}{2}\partial_\mu h \partial^\mu h    -\Bar{\psi}\gamma^\mu D_\mu\psi-U(h) - \frac{f}{\sqrt{2}}h\,\bar{\psi} \psi,
\end{align}
%
where the scalar potential is (see Fig.~\ref{fig:potential} for $\eta=0$)
\begin{equation}\label{higgs_potential}
    U(h) = \frac{\lambda}{16}\Big(h^2-v^2\Big)^2\,.
\end{equation}
Performing spontaneous symmetry breaking $h \rightarrow h + v$ in the Lagrangian~\eqref{theory_fund} gives rise to a well-defined scalar mass term,
\begin{equation}
    m_h^2 = \frac{\lambda v^2}{2}.
\end{equation}
However, for clarity of exposition, we prefer to work directly with~\eqref{theory_fund}, which has a simper analytical expression.
The Yukawa interaction is controlled by the coupling $f$, giving an effective mass to the fermion,
\begin{equation}
	m_\eff = \frac{f}{\sqrt{2}}\,h.
\end{equation}
It is also useful to define
\begin{equation}
    m_f = m_\eff(h = v) = \frac{f}{\sqrt{2}}\,v,
\end{equation}
which is the effective fermion mass when the scalar sits on the minimum $v > 0$. 

The fermionic field has a $U(1)$ global symmetry which ensures the conservation of the fermion number $N$ (the Noether charge). 

Our setup is similar to that of our previous work~\cite{DelGrosso:2023trq, DelGrosso:2023dmv}. An important difference is that here we relax the fine-tuning of the Yukawa coupling $f$ imposed in~\cite{DelGrosso:2023trq, DelGrosso:2023dmv}, which allows us to explore connections with more realistic particle-physics content, as later discussed.

We will consider spherically symmetric equilibrium configurations,
whose background metric can be expressed as
\begin{equation}\label{eq:general_spacetime}
\d s^2 = 
-e^{2u(\rho)} \d t^2 
+ e^{2v(\rho)}\d \rho^2 
+ \rho^2 (\d \theta^2 + \sin^2\theta \d \varphi^2),
\end{equation}
in terms of two real metric functions $u(\rho)$ and $v(\rho)$.

Fermions are treated through the Thomas-Fermi approximation~\cite{Lee:1986tr, DelGrosso:2023trq, DelGrosso:2023dmv}, practically meaning that they enter Einstein's equations as a perfect fluid characterized by an energy-momentum tensor of the form
\begin{equation}\label{eq:fermion_energy_momentum_tensor}
	T^{[f]}_{\mu\nu} = (W+P)u_\mu u_\nu + Pg_{\mu\nu},
\end{equation}
where $W$ is the energy density and $P$ is the pressure of the fluid, while they also enter the scalar field equation through the scalar density $S$. These quantities are defined as follows
\begin{align}
\label{fermion_energy}
W &= \frac{2}{(2\pi)^3} \int_0^{k_{\rm F}} \d^3 k \, \epsilon_k,
\\
\label{fermion_pressure}
 P &= \frac{2}{(2\pi)^3} \int_0^{k_{\rm F}} \d^3 k \hspace{0.1cm} \frac{k^2}{3\epsilon_k},
\\
\label{fermion_density}
S &=  \frac{2}{(2\pi)^3} \int_0^{k_{\rm F}} \d^3 k \hspace{0.1cm} \frac{m_\eff}{\epsilon_k}.
\end{align}
where $\epsilon_k = \sqrt{k^2 + m_\eff^2}$.
Notice that $W=W(x^\mu)$ through the spacetime dependence of $k_{\rm F}$ and $m_\eff$ (the same holds for $P$ and $S$). The integrals in Eq.~\eqref{fermion_energy},~\eqref{fermion_pressure},~\eqref{fermion_density} can be computed analytically as shown, for example, in Ref.~\cite{DelGrosso:2023trq}.

The fermion fluid is fully characterized once the Fermi momentum $k_\F$ is given.  Within the Thomas-Fermi approximation, it can be shown that
\begin{equation}
    k_{\rm F}^2(\rho) = \omega_{\rm F}^2e^{-2u(\rho)} - m_\eff^2\,,
\end{equation}
where $\omega_{\rm F}$ is the Fermi energy at the origin ($\rho=0$), which can be written in terms of the fermion central pressure $P(\rho = 0) \equiv P_c$ (see Ref.~\cite{DelGrosso:2023trq} for details).

In order to simplify the numerical integration, it is convenient 
to introduce the dimensionless quantities
\begin{equation}\label{dimensionless_variables}
x  = \frac{k_{\rm F}}{m_f}, \quad
y  = \frac{h}{v}, \quad
r = m_h\rho\,, \quad \tilde{\omega}_{\rm F} = \frac{\omega_{\rm F}}{m_f}\,,
\end{equation}
in terms of which the potential $U$ and kinetic $V=\unmezzo e^{-2v(\rho)} (\partial_\rho h)^2$ terms can be written as
\begin{align}\label{UV}
U 
\equiv  \mu^2 v_\F^2 \,\tilde{U}(y)\,,\qquad 
V 
\equiv \mu^2 v_\F^2 \,\tilde{V}(y).
\end{align}
Moreover, we introduce the following dimensionless fermionic quantities
\begin{equation}\label{dimensionless_fermionic_quantities}
\tilde{W} = \frac{W}{m_f^4}, \qquad
\tilde{P} = \frac{P}{m_f^4}, \qquad
\tilde{S} = \frac{S}{m_f^3} .  
\end{equation}

Finally, the field equations (i.e.\ the Einstein-Klein-Gordon equations with the addition of the Fermi momentum equation) take the compact form~\cite{DelGrosso:2023trq, DelGrosso:2023dmv}
\begin{align}\label{fund_sistema_dimensionlesse_dimensionlesse}
& e^{-2v}-1-2  e^{-2v} r\partial_r v = -\Lambda^2 r^2  \left [ \eta^4 \tilde{W} + \tilde{U} +  \tilde{V} \right],
\nn 
\\
&
e^{-2 v} - 1 + 2  e^{- 2v} r\partial_r u =\Lambda^2 r^2  \left [\eta^4 \tilde{P} - \tilde{U} +  \tilde{V}\ \right],
\nn 
\\
&
e^{-2v}\Big[  \partial_r ^2 y +  \Big(\partial_r u - \partial_r v + \frac{2}{r}\Big)\partial_r y \Big] -\frac{\partial \tilde{U}}{\partial y} - \eta^4\tilde{S} = 0 ,
		\nn 
		\\
& 
x^2 
= \tilde{\omega}_{\rm F}^2 e^{-2 u (r)} - y^2,
\end{align}
where $\tilde U$, $\tilde V$, $\tilde P$, $\tilde W$, and $\tilde S$ depend on $x$, $y$, and $r$.
The above equations contain two dimensionless combinations of the parameters\footnote{If we work in units such that $c = 1, \hbar \neq 1$, $\eta$ has dimensions of the square root of a coupling, i.e. $[\eta] = \hbar^{-1/4}$.}
\begin{align}\label{dimensionless_parameters}
	\Lambda &= \frac{\sqrt{8\pi}v}{m_p}, 
	\qquad 
	\eta = \frac{f}{(2 \lambda)^{1/4}} = \frac{m_f}{m_h^{1/2} v^{1/2}} \, ,
\end{align}
the physical interpretation of which will be discussed in the next section. Here $m_p=G^{-1/2}$ is the Planck mass.

NTSs in the model~\eqref{theory_fund} are static and spherically symmetric solutions to the above system of ordinary differential equations. The boundary conditions are the same as those used in Refs.~\cite{DelGrosso:2023trq, DelGrosso:2023dmv}, but in the present case\footnote{In the absence of fermions, there is also a solitonic solution that connects the two minima of the potential ($h = \pm v$). This configuration represents a domain wall (which is a topological soliton)~\cite{Saikawa:2017hiv}.} $y(r = 0) = \epsilon$, where $\epsilon$ is the initial
displacement with respect to the false vacuum $y = 0$.
The numerical procedure to build these solutions is the same as in Ref.~\cite{DelGrosso:2023trq}, to which we refer for more details.

Considering more than one Dirac fermion or including Majorana fermions in Eq.~\eqref{theory_fund} is a straightforward generalization. In the Thomas-Fermi approximation, this simply requires associating a Fermi momentum to each fermionic component and adding the relative energy density, pressure, and scalar density to the equations.

\section{Properties and interpretation}\label{sec_thestim}
Before presenting the numerical results, it is useful to discuss some general properties of these solutions.

The NTSs arising from Eq.~\eqref{theory_fund} are essentially fermion soliton stars with an effective positive cosmological constant inside, recently studied in~\cite{DelGrosso:2023dmv}. Therein, it was found that the energy of these solutions gets two contributions: one from the volume energy of the field and the other from a surface term. The relative importance of these two terms depends on the actual region of the parameter space.

In a field theory with spontaneous symmetry breaking, it is reasonable to expect $v \ll m_p$, which also guarantees that quantum gravity effects can be safely neglected. 
From Eq.~\eqref{dimensionless_parameters}, this implies $\Lambda \ll 1$, which corresponds to the regime where the energy is dominated by the volume contribution (see Eq.~(24) in Ref.~\cite{DelGrosso:2023dmv}). 
In this regime both the mass and the radius scale as $1/\Lambda$, while the parameter $\tilde{\omega}_{\rm F}$ scales as $1/\eta$. Moreover, the binding energy $E_B = M - m_f N$ of a configuration with $N$ trapped fermions scales as~\cite{DelGrosso:2023dmv}
\begin{equation}\label{eq:binding_energy}
    \frac{m_h E_B}{m_p^2} \sim \frac{1}{\Lambda}\Big(1-\eta\Big).
\end{equation}
Thus, for fixed $\Lambda\ll1$, the dimensionless binding energy is a linear decreasing function of $\eta$. Consequently, the bigger $\eta$ the more bound the configurations\footnote{This behavior could be modified by quantum corrections (see~\cite{Bagger:1991ph, Stewart:1996ez} for investigations along this line in the absence of gravity).}, with $\eta\approx1$ marking the transition from bound to unbound states. As we will discuss in detail later, for $\eta \lesssim 1$ the false vacuum pockets indeed disappear, and so does the nontrivial scalar profile of the ground state. Heuristically, this happens because the system has no benefit in investing some of its energy to build a false vacuum pocket to trap fermions that have a small mass gap between the two vacua of the scalar field. On the other hand, when $\eta \gtrsim 1$ the mass gap is big enough that it becomes energetically convenient to confine fermions in false vacuum pockets. For this reason, we will call the region \begin{equation}\label{eq:confining_regime?}
    \eta \gtrsim 1 \,,
\end{equation} 
\textit{confining regime}\footnote{In units such that $c = 1, \hbar \neq 1$, the perturbativity bound is $\hbar^{1/4}\eta \lesssim \sqrt{4\pi} \sim O(1)$. Restoring natural units, the condition~\eqref{eq:confining_regime?} corresponds to a strongly-coupled theory.} (in analogy to Refs.~\cite{DelGrosso:2023trq, DelGrosso:2023dmv}). Moreover, the latter condition ensures that the point $h = 0$ in Eq.~\eqref{higgs_potential} becomes a local minimum. Indeed, the effective scalar potential arising from Eq.~\eqref{theory_fund} is 
\begin{equation}\label{eq:effective_potential}
    \frac{U_\eff}{U_0} = \big(y^2-1^2\big)^2  + 8\,\eta^4 \tilde{S} \,y \, ,
\end{equation}
where $U_0 = \lambda v^4 / 16$. In order to analyze the behavior of the potential around the origin $y = 0$, we compute 
\begin{equation}
    \Big(\frac{U_\eff}{U_0}\Big)''(y = 0) = \frac{8 \eta ^4 \tilde{\omega}_\F^2}{\pi ^2}-4 \,.
\end{equation}
Thus, the point $y = 0$ will be a local minimum of the potential only if
\begin{equation}
    \Big(\frac{U_\eff}{U_0}\Big)''(y = 0) > 0\, ,
\end{equation}
which in turn yields  
\begin{equation}\label{eq:condition_omega_tilde}
    \tilde{\omega}_\F >  \frac{\pi }{\sqrt{2} \,   \eta^2} \,.
\end{equation}
Using the fact that $\tilde{\omega}_\F \sim 1/\eta$, one gets again Eq.~\eqref{eq:confining_regime?}. In Fig.~\ref{fig:potential} we explicitly show the behavior of the effective potential for different values of $\eta$ in the Newtonian regime, where the fermion scalar density $S$ is analytically expressed as a function of $y$ only.
%

It is important to stress that the NTSs discussed here are genuinely non-perturbative states. Indeed, in the limit in which one of the relevant couplings ($\lambda$ or $f$) is sent to zero, the energy diverges. When $\lambda\to0$, this happens because $\eta\to\infty$. When $f\to0$, instead, the fermion mass vanishes (and so does the mass gap) and consequently the hydrostatic equilibrium implies $h \to v$ (see Sec.~\ref{results_section}), as already anticipated from the previous heuristic argument. Therefore, NTSs arising from Eq.~\eqref{theory_fund} actually become standard fermion stars~\cite{Narain:2006kx} made of nearly massless fermions. In the zero-mass limit, the configuration does not have a finite radius and its energy diverges.
In both cases, it is impossible to describe these solutions perturbatively around $f\approx0$ or $\lambda\approx0$.

Notice that if the two vacuum states are degenerate, whenever $\Lambda\ll1$ there is no need for a strongly coupled fermion since the confining regime is achieved as long as $\eta \gtrsim \Lambda^{1/2}$ (see Ref.~\cite{DelGrosso:2023trq}), so $\eta$ can also be 
small. Indeed, in the latter case, the volume density of the scalar is zero, and the system needs much less energy to form a false vacuum pocket. 

Finally, we highlight that our semi-classical approach, in which the scalar field is described as a classical solution (while quantum effects are taken into account for fermions), is valid whenever the scalar self-interactions are weak, i.e. $\lambda \lesssim (4\pi)^2$.

\subsection{Numerical results}\label{results_section}

\begin{figure*}[t] 
    \centering
    \includegraphics[width=0.495\linewidth]{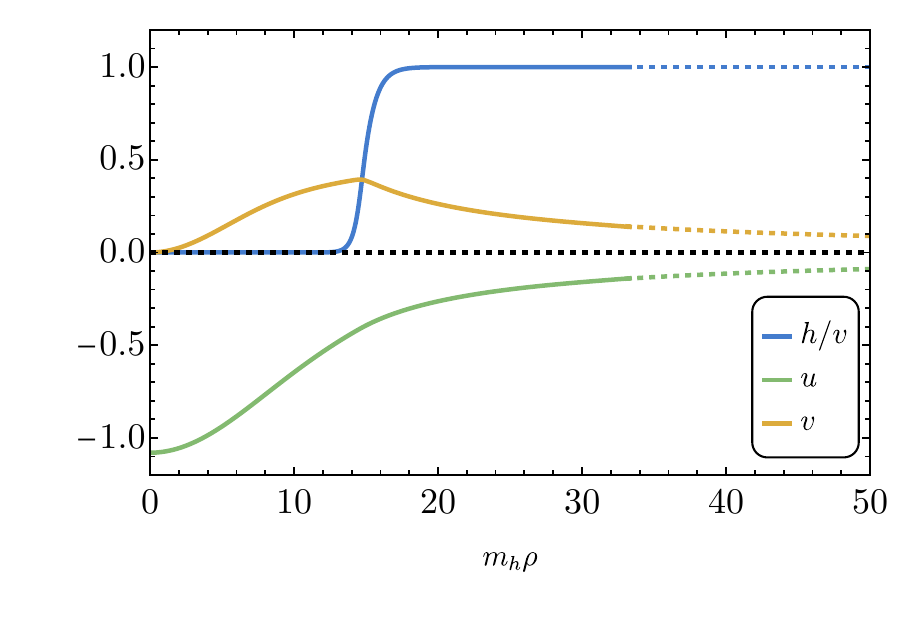}%
    \includegraphics[width=0.495\linewidth]{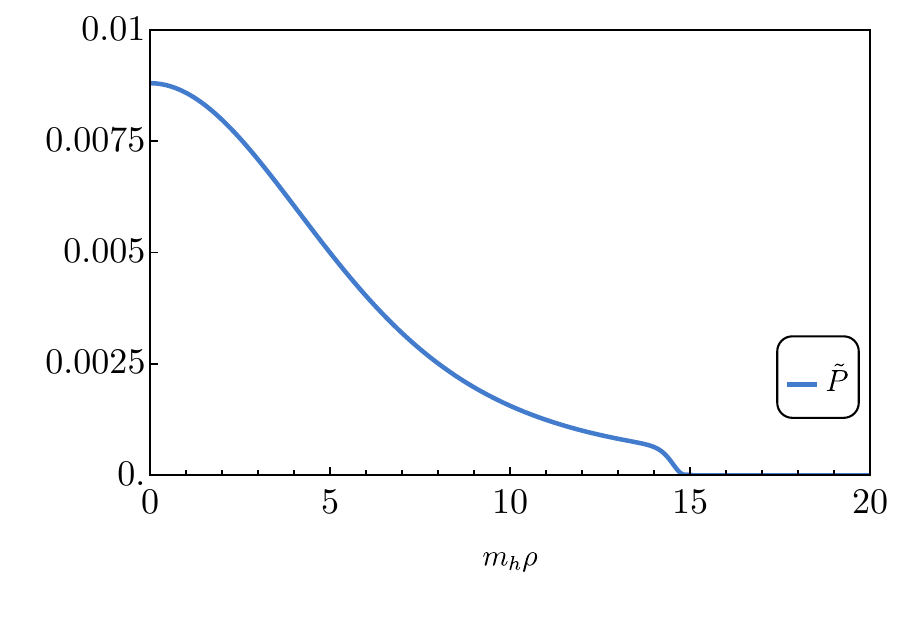}%
    \caption{Radial profiles of scalar field $h$, normalized with respect to $v$, metric functions $u$, $v$ (left panel) and fermion pressure (right panel) for a typical configuration ($\Lambda = 0.075$, $\eta = 4$). Continuous lines represent numerical data, whereas dashed lines reconstruct the asymptotic behavior of the solution by fitting with the Schwarzschild spacetime. The mass and radius of this configuration are $m_h M / m_p^2 \approx 4.08$ and $m_h R \approx 14.96$, the compactness is $C \approx 0.27$, while the solution parameters are $\tilde{P}_c = 8.8 \times 10^{-3}$ and $\log_{10}\epsilon = -19.00$. The binding energy is $m_h E_B / m_p^2 \approx -5.10$. 
    }
    \label{fig:example_solution}
\end{figure*}

In Fig.~\ref{fig:example_solution} we show an example of a solution with the radial
profiles for the metric, scalar field, and fermion pressure.

\begin{figure*}[t] 
    \centering
    \includegraphics[width=0.49\linewidth]{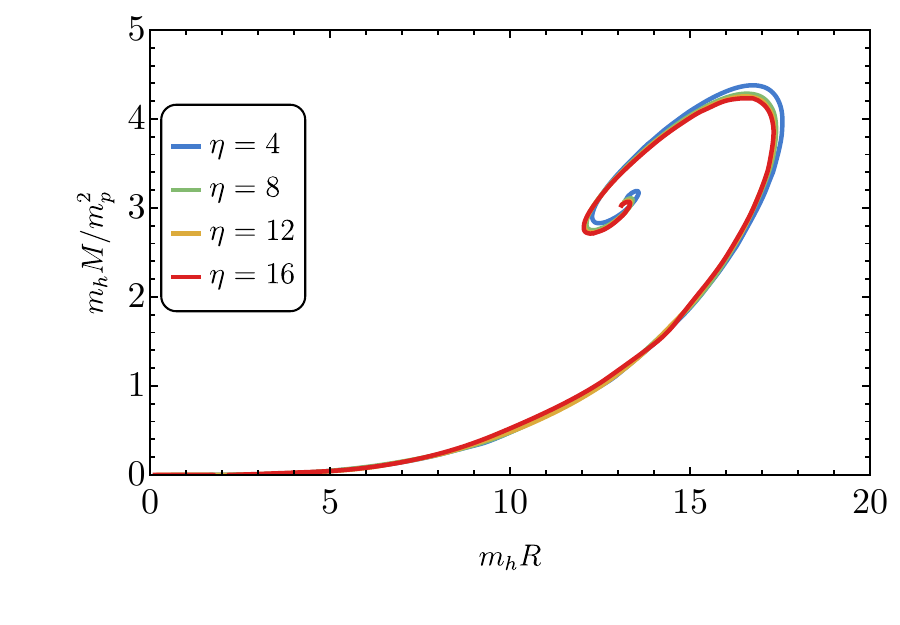}%
    \includegraphics[width=0.49\linewidth]{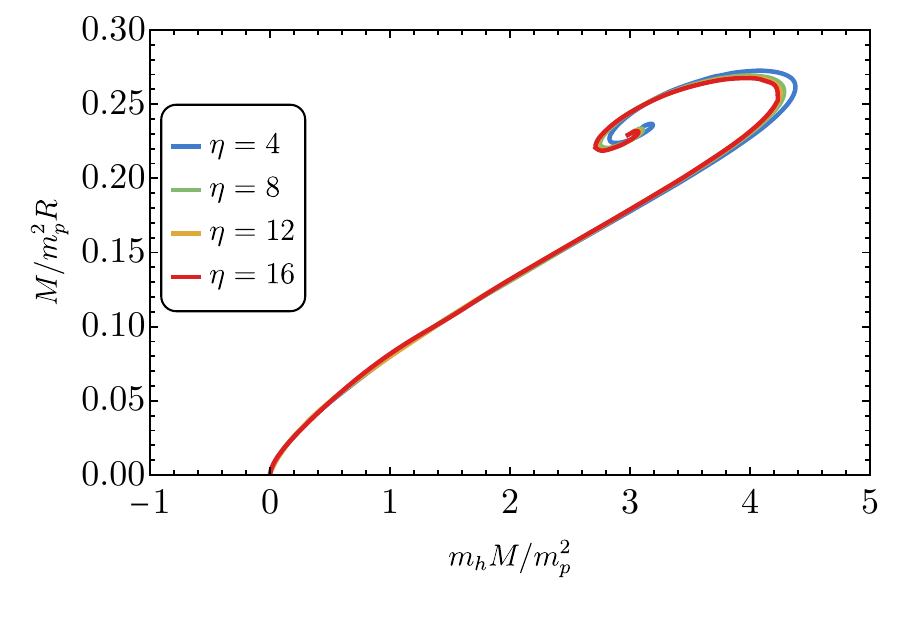}
    \caption{Mass-radius (left panel) and compactness-mass (right panel) diagrams for various values of $\eta$. We fixed $\Lambda = 0.075$. Varying $\Lambda$ does not change the results qualitatively. There exists a turning point in the $M$-$R$ diagrams at low masses, which cannot be seen from the figure, that proceeds towards the Newtonian limit of small $M$ and large $R$, similarly to what already shown in~\cite{DelGrosso:2023trq}.
    }
    \label{fig:massradious}
\end{figure*}

In Fig.~\ref{fig:massradious} we present the mass-radius and compactness-mass diagrams for various values of $\eta$ in the confining regime $\eta \gtrsim 1$. As anticipated, we observe that the mass-radius diagram is very weakly dependent on $\eta$.  
Moreover, the compactness becomes arbitrarily small along the tail of mass-radius diagram, giving rise to a Newtonian regime. 

\begin{figure*}[t] 
    \centering
    \includegraphics[width=0.49\linewidth]{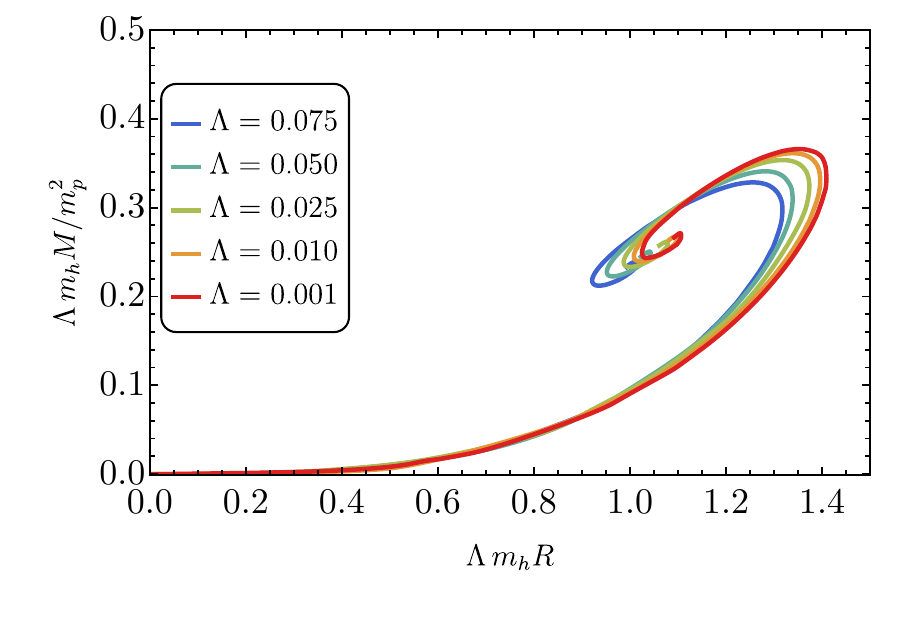}%
    \includegraphics[width=0.49\linewidth]{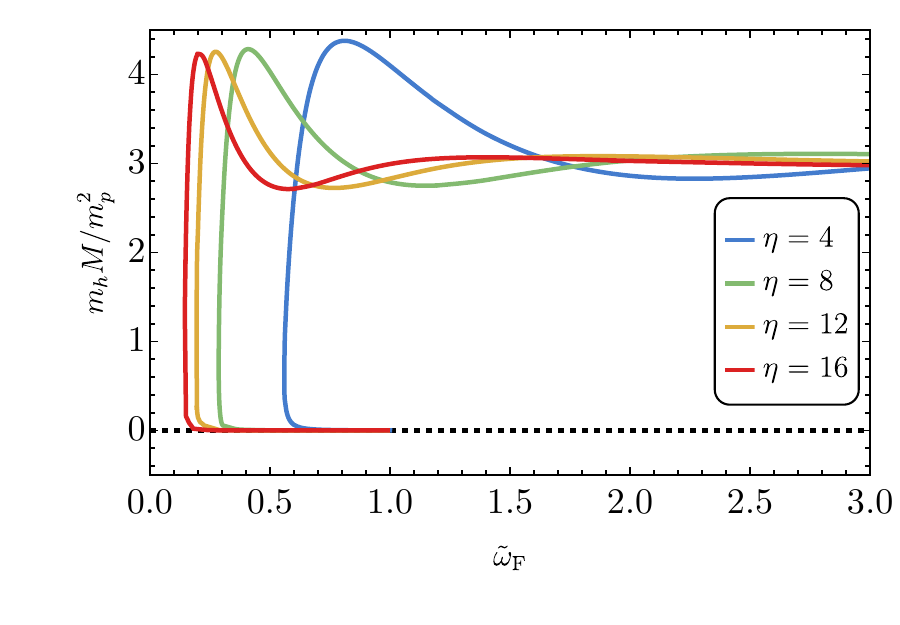}%
    \caption{
    \emph{Left panel:} Rescaled mass-radius diagrams for various $\Lambda$. We fix $\eta = 4$. Varying $\eta$ does not produce appreciable changes as long as we stay in the confining regime $\eta \gtrsim 1$. As expected, there is an approximate scaling law as $\sim 1/\Lambda$, which becomes more and more accurate as $\Lambda \to 0$.
    \emph{Right panel:} Mass as a function of $\tilde{\omega}_{\rm F}$ for various values of $\eta$. We fixed $\Lambda = 0.075$. Varying $\Lambda$ does not change the results qualitatively. Notice the existence of a minimum value for $\tilde{\omega}_{\rm F}$ at any given $\eta$.
    }
    \label{fig:rescaled_mass_radius}
\end{figure*}

In Fig.~\ref{fig:rescaled_mass_radius} (left panel), we show the rescaled mass $M\Lambda$ and rescaled radius $R\Lambda$, confirming the expected scaling in the $\Lambda\ll 1$ limit.
Overall, the mass-radius diagram is qualitatively similar to that of strange (quark) stars~\cite{Alcock:1986hz,Urbano:2018nrs}.

It is straightforward to identify a critical mass $M_c$ (and corresponding radius $R_c$) as the point of maximum mass in the $M$-$R$ diagram. As confirmed in Fig.~\ref{fig:rescaled_mass_radius}, for $\Lambda\lesssim 0.1$ those quantities behaves according to
\begin{align}
    \frac{m_h M_c}{m_p^2} = A(\eta)\, \frac{1}{\Lambda}, \qquad
    m_h R_c = B(\eta) \,\frac{1}{\Lambda}. 
\end{align}
As shown in Fig.~\ref{fig:massradious}, in the confining regime the dependence of the critical quantities on $\eta$ is very weak. Thus, $A$ and $B$ are approximately constant, and numerical fits show that $A \approx 0.36$ and $B\approx1.35$. 
We therefore obtain
\begin{align}\label{critical_quantities}
   M_c \approx M_\odot \Big(\frac{ 0.34\, {\rm GeV}}{q}\Big)^2, \quad
   R_c \approx 5.5 \,{\rm km} \Big(\frac{ 0.34\, {\rm GeV}}{q}\Big)^2\,,
\end{align}
where $q = (m_h v)^{1/2}$, in terms of which the condition~\eqref{eq:confining_regime?} can be written as
\begin{equation}\label{critical_condition}
    m_f \gtrsim q.
\end{equation}
Finally, the compactness of the critical configuration is $GM_c/R_c\sim A/B \sim 0.27$, slightly higher than that of a typical neutron star.

In Fig.~\ref{fig:rescaled_mass_radius} (right panel) we plot the mass as a function of $\tilde{\omega}_{\rm F}$ for various values of $\eta$ at fixed $\Lambda = 0.075$, in the confining regime. In the low-mass region of these curves, where $m_h M / m_p^2 \approx 0$, $\tilde{\omega}_{\rm F}\to 1$. In that limit $\epsilon \to 1$, the fluid energy density (together with its number density) vanishes and thus the false vacuum pocket is lost. As long as we depart from the $m_h M / m_p^2 \approx 0$ branch, $\epsilon$ becomes a small number and the false vacuum pocket is recovered. From Fig.~\ref{fig:rescaled_mass_radius} (right panel) we also notice that, for any given $\eta$, $\tilde{\omega}_{\rm F}$ cannot be arbitrarily small. This means that configurations in which $\epsilon$ is small (i.e. truly false vacuum pockets) exist only above a minimum value $\tilde{\omega}_{\rm F}^{\rm min}$. We numerically verified that the scaling for $\tilde{\omega}_{\rm F}^{\rm min}$ is
\begin{equation}
    \tilde{\omega}_{\rm F}^{\rm min} \approx \frac{2}{\eta}\, .
\end{equation}
Notice that this quantity is independent of $\Lambda$. Since the number density $n$ in the core is proportional to $\omega_{\rm F}^3$ (see Sec.~II~C in Ref,~\cite{DelGrosso:2023dmv} for further details), we estimate 
\begin{align}\label{eq:minimum_density}
    n^{\rm min} &\approx \frac{1}{3\pi^2} m_f^3  \,(\tilde{\omega}_{\rm F}^{\rm min})^3 \approx 
    \frac{8} {3\pi^2}(m_h v)^{3/2}\, .
\end{align}
The latter quantity can be interpreted as the minimum Noether charge per unit volume that allows for the existence of false vacuum pockets in the model~\eqref{theory_fund}. The configuration with the minimum possible Noether charge density has some peculiar properties. Calling $M_{\rm min}$ and $R_{\rm min}$, respectively, its mass and radius, numerical analysis shows that 
\begin{align}\label{eq_minimal_quantities}
    \frac{m_h M_{\rm min}}{m_p^2} \approx 1.00, \qquad
    m_h R_{\rm min} \approx \frac{2.24}{\Lambda^{2/3}}. 
\end{align}
Hence, in the $\Lambda \ll 1$ limit this special configuration is pushed towards the tail of the mass-radius diagram, where the compactness $G\,M_{\rm min}/R_{\rm min} \sim \Lambda^{2/3} \to 0$. We conclude that NTSs arising from Eq.~\eqref{theory_fund} scan a range of masses and radii starting from the configuration in Eq.~\eqref{eq_minimal_quantities} deep in the tail of the mass-radius diagram, where relativistic gravitational effects are negligible, up to the critical configuration in Eq.~\eqref{critical_quantities}, above which the NTS is expected to undergo gravitational collapse.

\begin{figure*}[t] 
    \centering
     \includegraphics[width=0.49\linewidth]{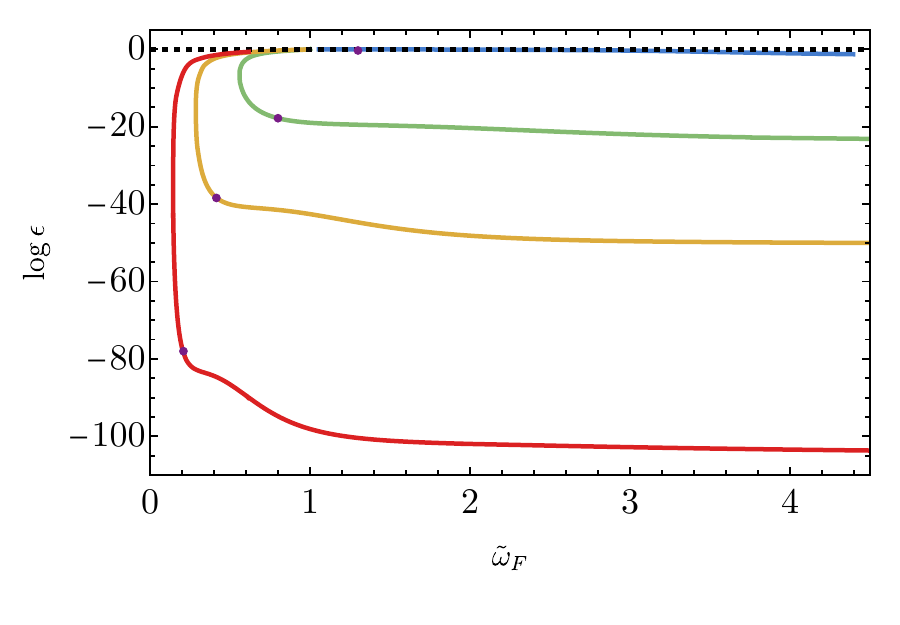}%
     \includegraphics[width=0.49\linewidth]{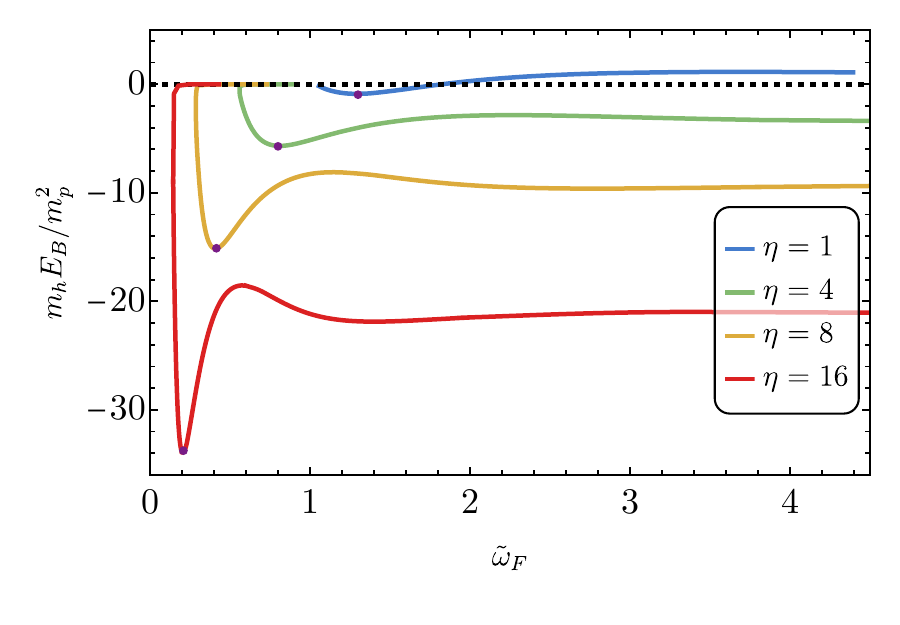}%
    \caption{The initial displacement (left panel) and binding energy (right panel) for various values of $\eta$. We fixed $\Lambda = 0.075$. Varying $\Lambda$ does not change the results qualitatively. As expected, the bigger $\eta$, the lower the binding energy. The configurations corresponding to the critical ones are highlighted with a violet marker. 
    }
    \label{fig:bindingenergy}
\end{figure*}

In Fig.~\ref{fig:bindingenergy}, we show the initial displacement $\epsilon$ and the binding energy of the configurations as a function of $\eta$. As expected from the discussion in Sec.~\ref{sec_thestim}, the bigger is $\eta$ the lower is the binding energy. In particular, we observe that the binding energy of the critical configurations, marked with a violet dot in Fig.~\ref{fig:bindingenergy}, decreases linearly in $\eta$, as expected. Around $\eta \approx 1$, we find configurations with positive binding energy, in agreement with the estimate in Eq.~\eqref{eq:binding_energy}. 

From the left panel of Fig.~\ref{fig:bindingenergy}, we observe that the bigger is $\eta$, the lower is the displacement $\epsilon$, whereas, for $\eta \approx 1$, the displacement is $\epsilon \approx 1$, meaning that the scalar field is practically already on the true vacuum. In the latter case, as anticipated from the discussion in Sec.~\ref{sec_thestim}, the effective fermion mass is already very near to $m_f$ and the picture of the configuration as a false vacuum pocket is lost.  
Therefore, we confirm that the ground state acquires a non-trivial scalar profile only if $\eta$ is above a particular threshold. The numerical analysis done within the Thomas-Fermi approximation shows that already with $\eta \lesssim 2$ we enter the deconfining regime (with the appearance of several turning
points in both the mass and the radius, 
similarly to what is shown in Ref.~\cite{DelGrosso:2023trq}) and for $\eta = 1$ all the configurations with $E_B < 0$ have $h \approx v$.

\subsection{Stability of NTSs}\label{standard_stability_section}

There are three main independent mechanisms through which a NTS may decay in time. First of all, if the binding energy is positive, a configuration with $N$ fermions trapped inside is triggered to disperse into free $N$ particles by quantum and/or thermal fluctuations~\cite{Lee:1991ax}. As discussed in Sec.~\ref{results_section}, as long as Eq.~\eqref{critical_condition} is satisfied, the configurations are bound.

Second, NTSs may be unstable under classical perturbations. A full investigation of the latter point goes beyond the scope of this work. However, in the confining regime, the shape of the mass-radius diagram is qualitatively similar to that of strange (quark) stars~\cite{Alcock:1986hz,Urbano:2018nrs}. Therefore, we expect the configurations below the critical mass to be at least radially stable, as it happens for ordinary stars~\cite{Shapiro:1983du}.

Third, the configurations could be unstable under fission into smaller configurations. Calling $E(N)$ the energy of a NTS with $N$ trapped particles inside, whenever
\begin{equation}\label{eq:fission}
    E(N_1) + E(N_2) > E(N_1 + N_2)
\end{equation}
fission is forbidden. The above equation can be recast in the following condition (also known as the Vakhitov–Kolokolov stability criterion~\cite{Friedberg:1976me, Almumin:2021gax})
\begin{equation}\label{eq:fission2}
    \frac{d^2 E}{dN^2} < 0\,.
\end{equation}
For the configurations laying below the critical mass, we numerically checked that Eq.~\eqref{eq:fission2} holds.

\section{Applications of the model}
In this section, we discuss explicit realizations of the model under investigation, embedded in various theories.

\subsection{Neutron soliton stars}
A simple application of non-perturbative vacuum scalarization is in the context of the linear sigma model~\cite{Machleidt:2011zz}, a low-energy effective theory of QCD that implements spontaneous chiral symmetry breaking. In this context, the fermion in Eq.~\eqref{theory_fund} is just a nucleon, while the scalar field is a scalar meson called $\sigma$. 
Using the benchmark values $m_h \sim 500\,{\rm MeV}$~\cite{Morones-Ibarra:2006and}, $v \sim f_\pi \sim 130 \,{\rm MeV}$ in Eq.~\eqref{eq_minimal_quantities}, the minimal configuration is estimated to be
\begin{align}\label{}
 M_{\rm min} \approx 10^{-19} \,M_\odot, \qquad 
   R_{\rm min} \approx 1\, {\rm cm},
\end{align}
while the critical one in Eq.~\eqref{critical_quantities} is
\begin{align}\label{benchmark_values_NS}
   M_c \approx 2 \,M_\odot, \qquad 
   R_c \approx 10\, {\rm km},
\end{align}
with a compactness of $M_c/R_c\sim 0.27$, that is
slightly larger than that of an ordinary neutron star.
Moreover, being the mass of the nucleon $m_f \sim 1 \,{\rm GeV}$, we get $\eta \approx 4$, ensuring that the configurations lay in the confining regime. These estimates are in agreement with the ones given in Ref.~\cite{Witten:1984rs}.

Remarkably, the heaviest neutron stars discovered~\cite{Romani:2022jhd} have a mass compatible with the estimate in Eq.~\eqref{benchmark_values_NS}. 
In the absence of a scalarization mechanism, it is well known that a standard degenerate Fermi gas of neutrons can support compact stars only up to $\sim 0.7 M_{\odot}$~\cite{Oppenheimer:1939ne}, and nuclear interactions are therefore needed to explain heavy neutron stars.
In our model, instead, we can achieve the same result by simply coupling the degenerate Fermi gas to a scalar field.

We are neglecting the energy density of the pions assuming that their mean values on the ground state are zero. However, they are inevitably present in the linear sigma model and the inclusion of a pion condensate could lead to interesting phenomenology~\cite{Vijayan:2023qrt}. Exploring these effects, however, would require a more involved analysis.

Our simple estimate suggests that the ground state of the heaviest neutron stars can in fact be scalarized and their core is made of a false vacuum pocket where the chiral symmetry is unbroken~\cite{Balkin:2023xtr, Buballa:2015awa}. Such NTSs may be formed in the early universe, through the mechanisms mentioned in the introduction, or during the merger of two neutron stars, if the energy involved is high enough ($\sim f_\pi$) to allow for the formation of false vacuum pockets and consequent trapping of nucleons.

\subsection{Higgs false vacuum pockets}

Non-perturbative vacuum scalarization may play a role also in the Higgs sector of the Standard Model~(SM). Indeed, neglecting gauge interactions, Eq.~\eqref{theory_fund} naturally arises after electroweak symmetry breaking, as shown in Appendix~\ref{SM_derivation}. In this scenario, $h$ is the physical Higgs field and $\psi$ a SM fermion. Using $m_h \sim 125 \, {\rm GeV}$ and $v \sim 246 \, {\rm GeV}$, the minimal configuration is found to be
\begin{align}\label{benchmark_valuesMIN}
   M_{\rm min} &\approx  10^{-21} \,M_\odot, \qquad
   R_{\rm min} \approx 0.2\, {\rm \mu m},
\end{align}
while the critical mass and radius are 
\begin{align}\label{benchmark_values}
   M_c &\approx (4 \times 10^{-6}) \,M_\odot, \qquad 
   R_c \approx 2\, {\rm cm}.
\end{align}
These configurations are compelling candidates for non-particle dark matter and exotic compact objects. Indeed, it is intriguing to note that the critical mass range is naturally compatible with the OGLE+HSC observation of some microlensing events~\cite{Niikura:2017zjd, Niikura:2019kqi}.
However, non-perturbative vacuum scalarization occurs only if there is a strongly coupled fermion to the Higgs boson, that is, only if the condition in Eq.~\eqref{eq:confining_regime?} is satisfied. For the Higgs parameters, Eq.~\eqref{eq:confining_regime?} gives $m_f \gtrsim 175 \, {\rm GeV}$, which is very near the top quark mass $m_{\rm top}$. Nevertheless, as discussed in Sec.~\ref{results_section}, numerical computations show that we need $\eta \gtrsim 2$ to form bound false vacuum pockets, meaning that $m_f$ should be at least $\sim 350 \, {\rm GeV}$. This suggests that non-perturbative vacuum scalarization does not occur in the SM\footnote{A similar conclusion is found in Ref.~\cite{Dimopoulos:1990at}, where however the authors look for Higgs vacuum deformation around just one top quark (and not a collection thereof). Moreover, they neglect gauge interactions (and also gravity, since they are interested in microscopic states).}. In addiction, even if the quark top mass was above the threshold, the electric charge carried by the fermions would produce a repulsive Coulomb force that in general renders large configurations unstable~\cite{Lee:1988ag, Gulamov:2015fya}. Indeed, the total electric charge is estimated to be $Q \sim (2/3 \, e) \, n \, R^3$. For the minimal configuration this gives $Q_{\rm min} \sim 10^{14} \, {\rm C}$, whereas for the critical one $Q_{c} \sim 10^{29} \, {\rm C}$. In both cases, the total electric charge is way above the maximum charge compatible with hydrostatic equilibrium~\cite{Fabbrichesi:2019ema}, i.e. $10^{-22} \, {\rm C}$ for the minimal configuration, and $10^{-7} \, {\rm C}$ for the critical one.

However, top quarks also carry color charges. This gives a possible caveat to the previous arguments and leaves open the possibility of forming Higgs false vacuum pockets using only SM fields. To support the latter hypothesis, it is enough to estimate the fermion number density $n$ inside the object. Using Eq.~\eqref{eq:minimum_density}, we obtain
\begin{equation}
    n\sim m_f^3 \tilde{\omega}_\F^3 \sim \frac{m_f^3}{\eta^3} \sim (m_h v)^{3/2}.
\end{equation}
For the Higgs field parameters, $n \sim 7 \times 10^{47} {\rm \, cm^{-3}}$, which is orders of magnitude larger than the nuclear matter density, $n_{\rm nucl} \sim 10^{38} {\rm \, cm^{-3}}$. Such high densities are expected to give rise to exotic states, in which interactions among fermions cannot be anymore neglected, such as a top/anti-top bound state~\cite{Froggatt:2014tua} or a colored superconductor~\cite{Alford:2007xm}. These scenarios could allow for bulk matter neutrality with respect to both electric and color charges. A self-consistent computation where Eq.~\eqref{eq:fermion_energy_momentum_tensor} is substituted by a new energy-momentum tensor, in which all these effects are taken into account, may lower the aforementioned threshold and allow for the existence of bound configurations.

Non-perturbative vacuum scalarization may naturally occur in extensions of the SM by a fourth generation of (heavy) fermions. However, the simplest models along this direction are ruled out~\cite{Lenz:2013iha}. A viable possibility is adding only a fourth family of chiral leptons strongly coupled to the Higgs sector, without modifying the quark content. This would allow for evading bounds on the number of generations coming from the Higgs decay into gluons.

It is important to stress that, regardless of the particular beyond-SM context in which Eq.~\eqref{theory_fund} is obtained, as long as $h$ is the Higgs boson, there are no free parameters in the model, being the mass and radius of the configurations very weakly dependent on $m_f$.

\subsubsection{False vacuum pockets evaporation}
Beyond the three standard mechanisms mentioned in Sec.~\ref{standard_stability_section}, there could be other ways of destroying the object, depending on the specific embedding of the model.
Here we mostly focus on the case of (compact) Higgs false-vacuum pockets, but the argument can be extended also to other models in which the scalar $h$ has Higgs-like couplings.

Let us assume that the fermions interact with (asymptotically) lighter SM particles (other than the Higgs boson). Inside the pocket, being the fermions effectively massless, their lifetime is expected to be extremely long. However, on the boundary of the pocket, the fermions re-acquire their mass and decays into lighter SM particles become kinetically allowed. Under the assumption that all the decay products leave the pocket (effectively subtracting energy from the configuration), it is possible to give a rough estimate of the lifetime. The number of fermions in the boundary is $N^f_{\rm shell} \sim m_h^{-1} R^2 \omega_\F^3$, where $m_h^{-1}$ is the size of the region where the fermions acquire their mass and $\omega_\F^3$ is the fermion number density. The number of fermions that leave the object per unit time is estimated as 
\begin{equation}
   \frac{d N^f_{\rm shell}}{dt} \sim m_h^{-1} R^2 \omega_\F^3 \times \Gamma(m_f \rightarrow \text{ something}).
\end{equation}
A naive guess is $\Gamma(m_f \rightarrow \text{ something}) \sim m_f$. Moreover, each fermion decay subtracts from the object an energy $\sim m_f$. Therefore, the total rate of energy loss is
\begin{equation}
    \frac{d E_{\rm loss}}{dt}\sim m_f\frac{d N^f_{\rm shell}}{dt} \sim m_h^{-1} R^2 \omega_\F^3 m_f^2.
\end{equation}
The lifetime is estimated as 
\begin{equation}
    t_{\rm decay} \sim\frac{M}{d E_{\rm loss}/dt} \sim \frac{M}{m_h^{-1} R^2 \omega_\F^3 m_f^2}
\end{equation}
Since $M \sim m_p^2 / m_h \Lambda$, $R \sim 1/m_h \Lambda$, $\omega_\F = m_f \tilde{\omega}_\F \sim m_f / \eta$, one finally gets
\begin{equation}\label{key}
	t_{\rm decay} \sim \frac{m_p^2 m_h^2\, \Lambda \,\eta^3}{m_f^5} \times \frac{v^2}{v^2} = \frac{1}{\Lambda \eta} \frac{1}{m_f} \sim \frac{\tilde{N}^{1/3}}{m_f}
\end{equation}
We notice that in the limit $N \to \infty$ the lifetime would be infinity.
Now,
\begin{equation}\label{key2}
	t_{\rm decay} \sim  \frac{1}{\Lambda \eta} \frac{1}{\eta} \frac{1}{\sqrt{m_h v}} \sim \frac{1}{\Lambda\, \eta^2} \,10^{-27}{\,\rm sec}
\end{equation}
Using the Higgs parameters, one gets $\Lambda \sim 10^{-16}$ and
\begin{equation}\label{decay_shell}
	t_{\rm decay} \sim \frac{10^{-11}{\, \rm sec}}{\eta^2}.
\end{equation}
The actual computation of the lifetime should be carried out using a background field method (see e.g. Ref.~\cite{Cohen:1986ct}) and is complicated by the fact that we are in a regime of strong gravitational field\footnote{We expect that strong gravity can only increase the lifetime of the object.}. 

Moreover, there are several subtleties that can drastically change the latter estimate. First of all, our computation relies on the assumption that the fermions are a gas of weakly interacting particles. Conversely, if one adds sizeable interactions between fermionic particles, the lifetime may become much bigger, just like the lifetime of the neutron inside a nucleus could be bigger than the age of the Universe, whereas the lifetime of a free neutron is very short. Second, in a beyond-SM scenario, the decay into lighter SM particles can be forbidden if the fermions in the configuration possess extra symmetries (see e.g.~\cite{Bao:2013zua}) or do not couple with other SM particles.

What can not be disentangled from the other SM content is the Higgs field. In principle, some Higgs quanta of the system may be converted into other SM particles, producing a net particle flux. Since the Higgs boson is in a non-linear wave configuration, computing the latter quantity is a non-trivial problem. 

We give an estimate through the following argument. A NTS can be quantum mechanically described as a superposition of many coherent states (see e.g.~\cite{Dvali:2015jxa}). Each soliton quantum represents an interacting state which has in principle nothing to do with the standard perturbative states of the theory. The soliton size $R$ is set by the energy $\mu$ of these soliton quanta and the configuration is dominated by modes with momentum $k \lesssim \mu$. In particular,
\begin{equation}\label{key3}
	R \sim \frac{1}{\mu}\,.
\end{equation}
Since in our model $R \sim \frac{1}{m_h \Lambda}$, it is natural to estimate $\mu \sim m_h \Lambda \sim 10^{-14} \, {\rm GeV} = 10^{-5} \, {\rm eV}$. Therefore, just by energy conservation, the soliton quanta are not able to produce asymptotically a SM (massive) particle. %
Even neutrino (asymptotically) production is forbidden since $m_{\rm neutrino} \sim  10^{-2} \, {\rm eV} \gg \mu$.

What we cannot exclude is asymptotically photon production. However, since the Higgs boson does not couple to photons at the tree level, the overall effect is suppressed by loop effects. A rough estimate for the number of photons produced $N_\gamma$ gives 
\begin{equation}
    \frac{dN_\gamma}{dt} \sim N_{\mu}\times\Gamma(\mu \rightarrow \gamma \gamma) \sim \frac{1}{\Lambda^4} \times  \,\alpha^2 \mu 
\end{equation}
where $\alpha$ is the fine-structure constant and $N_\mu$ is the number of soliton quanta which is estimated to be $\sim m_h^2 v^2 R^3 / \mu \sim 1/\Lambda^4$. Since the energy of each photon is $\sim \mu$, the change in the total energy is roughly $\mu \,dN_\gamma/dt$, which in turn gives the time needed to destroy the object 
\begin{equation}\label{lifetime_photons}
	\frac{M}{\mu\, dN_\gamma/dt} \sim \frac{1}{\Lambda} \frac{1}{\alpha^2 m_h} \sim 10^{-6}\,{\rm sec},
\end{equation} 

An important caveat of Eq.~\eqref{lifetime_photons} is that we are estimating the matrix element $\Gamma(\mu \rightarrow \gamma \gamma)$ as if the soliton quanta were the standard perturbative Higgs quanta of mass $\mu$. However, we have already highlighted that the soliton quanta are purely interacting states which have nothing to do with the standard perturbative states. Therefore, a proper computation is needed to reliably estimate $\Gamma(\mu \rightarrow \gamma \gamma)$. 

\subsubsection{Accretion-driven collapse in the early universe?}
Even in those scenarios where the solitons evaporate (as in the case of the false-vacuum Higgs solitons), they might still play a role during the evolution of the Universe, for example, if they survive long enough to accrete a sizeable fraction of their mass.

Indeed, while for purely scalar solitons (such as boson stars) scalar accretion-driven 
collapse to a black hole is prevented by gravitational cooling~\cite{Seidel:1993zk,Liebling:2012fv,Brito:2015yga}, our solutions have a fermionic core which can undergo gravitational collapse upon accretion of ordinary matter.

In the simplest scenario, Higgs balls are produced in the radiation-dominated era with an initial mass $M_i$. In principle, such objects are able to accrete significantly before evaporation and one can therefore estimate the time needed to reach the critical mass $M_c$ above which the object can collapse into a (primordial) black hole.
Following Refs.~\cite{Zeldovich:1967lct, Carr:1974nx}, we estimate the mass $M$ of the object at the time $t$ as
\begin{equation}\label{keyG}
	GM = \frac{t}{1+\frac{t}{t_i} \Big(\frac{t_i}{G M_i} -1\Big)} \, ,
\end{equation}
where $t_i$ is the time of soliton formation and $t_i\gtrsim G M_i$ since $M_i$ must be smaller than the horizon mass $t_i/G$.
At $t\gg t_i$, the final mass asymptotically is
\begin{equation}
    M\to \frac{M_i}{1- G M_i/t_i}\,,
\end{equation}
and can be significantly higher than $M_i$ if $GM_i\sim t_i$.

Assuming that the object decays in a time $t_d$ and requiring $M(t_d)>M_c$ for black-hole formation, we obtain
\begin{equation}
    M_i> \frac{M_c}{1+ G M_c\left(\frac{1}{t_i}-\frac{1}{t_d}\right)} \sim \frac{M_c}{1+ \frac{G M_c}{t_i}}\,,
\end{equation}
where the last step is valid if $t_d\gg t_i$.

The above condition might or might not hold in specific models. For the case of the Higgs ball, we assume  
$M_c\sim 4\times 10^{-6} M_\odot$, $t_d=t_i+t_{\rm decay}$, and formation at the electroweak phase transition ($t_i\approx 2\times10^{-11}\,{\rm sec}$). Therefore, for $\eta\approx1$, these objects will collapse to black holes before evaporation only if 
\begin{equation}\label{collapse}
	M_i \gtrsim 3\times 10^{-6} M_\odot\,,
\end{equation}
which is slightly smaller than the maximum mass and than the horizon mass at the time of formation. 

\subsection{Dark soliton stars}
The simplest beyond-SM scenario where non-perturbative vacuum scalarization is realized is in the context of the dark sector paradigm~\cite{Alexander:2016aln}, where we interpret Eq.~\eqref{theory_fund} as embedded in the dark sector.
%
The SM Higgs field $h_{\rm SM}$ interacts with the dark scalar $h$ through the unavoidable scalar portal $\propto h_{\rm SM}^2 h^2, h_{\rm SM} h^2, h_{\rm SM}^2 h$. We assume $ m_h \gtrsim m_{h_{\rm SM}}$, in order to kinematically forbid the Higgs boson direct decay into the dark scalar, allowing for evading collider constraints on the listed couplings.

In this framework, the evaporation bounds discussed in the previous section can be easily evaded. Indeed, assuming that $\psi$ is the lightest fermion in the dark sector, decays on the boundaries are forbidden. Moreover, remembering that NTSs arising from Eq.~\eqref{theory_fund} are made of scalar quanta with characteristic energy $\mu \sim m_h \Lambda$, as long as
\begin{equation}\label{eq:condition_on_dark_scalar1}
    m_h \Lambda < m_{h_{\rm SM}} \sim 125  \, {\rm GeV} \,,
\end{equation}
the soliton quanta cannot produce asymptotically SM Higgs particles. Combining the latter equation with the requirement $ m_h \gtrsim m_{h_{\rm SM}}$, we finally get
\begin{equation}\label{eq:condition_on_dark_scalar2}
\Lambda  < \frac{m_{h_{\rm SM}}}{m_h} \lesssim 1\,.
\end{equation}
Since it is natural to assume $\Lambda \ll 1$, as discussed in Sec.~\ref{sec_thestim}, Eq~\eqref{eq:condition_on_dark_scalar2} is easily satisfied even for a dark scalar much heavier than the SM Higgs boson. 

If the scalar quartic $\lambda$ is an $O(1)$ number, the condition $m_h \gtrsim m_{h_{\rm SM}}$ translates into $q \gtrsim q_{\rm SM}  \approx 175 \, {\rm GeV}$. Therefore, we conclude that these dark soliton stars are expected to be stable and support compact objects of masses naturally in the subsolar regime, according to Eq.~\eqref{critical_quantities}. In particular, for $m_h \sim v \sim 10^2 \, {\rm GeV}$, we have a stable configuration with parameters similar to Eq.~\eqref{benchmark_values}.

The dynamical formation of NTSs arising in a similar framework has been recently studied in~\cite{Hong:2020est}, through a first-order cosmological phase transition. The configurations produced, dubbed by the authors \textit{Fermi balls}, are the non-relativistic limit of our solutions. If an initial distribution of Fermi balls is able to accrete and cool down, the final state will be a scalarized ground state (see \textbf{(c)} in Fig.~\ref{fig:illustration}), well-described by our solutions.

\section{Conclusions}\label{sec:conclusions}
In this work, we outlined the non-perturbative vacuum scalarization as a general mechanism to support new solitonic objects that can form in the early universe and serve as dark matter candidates. For concreteness, we considered a theory with a real scalar field coupled to a fermion in the context of General Relativity and found solutions describing self-gravitating compact objects where the scalar field acquires a non-trivial profile due to non-perturbative effects.

Besides the specific examples discussed in this work, other scenarios where this non-perturbative vacuum scalarization mechanism can play a role are fourth-generation models with extended Higgs sector~\cite{Luty:1990bg}, asymmetric dark matter models~\cite{Petraki:2013wwa}, mirror symmetries, minimal supersymmetric SM~\cite{Csaki:1996ks,Delgado:2020url}, Type II see-saw mechanism~\cite{Antusch:2018svb}, grand unified theories, inflation and cosmological phase transitions~\cite{Carena:2004ha,Angelescu:2018dkk}. Moreover, the inclusion of gauge fields into the solutions could give rise the important effects (see e.g. Ref.~\cite{Lee:1988ag, Gulamov:2015fya, Endo:2022uhe}), which we plan to explore in future work.

Although NTSs are energetically favored and stable under (at least) radial perturbations, couplings to other fields or self-interactions might induce evaporation of the solution, the details of which depend on the specific embedding of the model in a given theory. Future work should assess the role of fermion interactions for the evaporation time scale in given models. Even in cases in which this time scale is short relative to typical astrophysical scales, in the early universe NTSs might have enough time to accrete past the maximum mass or merge with other objects, with the possibility of forming primordial black holes in both cases~\cite{Khlopov:1985jw, Kawana:2021tde, DeLuca:2021pls, Kim:2023ixo, Flores:2023zpf, Lewicki:2023mik, Huang:2022him}. Interestingly, this scenario would produce primordial black holes with a mass fixed in terms of the maximum mass of the soliton, regardless of the formation epoch. As we have shown, certain realizations of our model naturally lead to compact objects in a mass range that is compatible with the OGLE+HSC anomaly for some microlensing events~\cite{Niikura:2017zjd, Niikura:2019kqi}.

A further important extension concerns the dynamical formation of these solutions. Many different channels have already been proposed in the literature. For example, NTSs arising from Eq.~\eqref{theory_fund} can be produced by a first-order cosmological phase transition~\cite{Hong:2020est}. Alternatively, the Yukawa interaction, if enough long-range, drives clustering and leads to the formation of compact NTSs~\cite{Domenech:2021uyx, Domenech:2023afs, Flores:2023zpf}.
Further possible formation channels, worth exploring in future work, are the following. First of all, statistical fluctuations inevitably present even during a crossover, are in principle able to provide a large concentration of fermions~\cite{Griest:1989cb} and can be, therefore, the dominant source of charge fluctuations which leads to NTS formation through solitosynthesis. Moreover, one could consider a configuration where there is a gas of $N$ free (massive) fermions moving in the true scalar vacuum $h \equiv v$. If $\eta\gtrsim 1$ and for a sufficiently large perturbation of the scalar field with energy above a certain threshold, we expect to end up in the true (scalarized) ground state.

The idea that a system can scalarize non-linearly, i.e. only if perturbed above a certain threshold, has been already numerically studied for scalar perturbations around a Schwarzschild black hole in scalar-Gauss-Bonnet gravity theories~\cite{Doneva:2021tvn}. An interesting extension of our work would be performing similar simulations in our model, possibly within a cosmological scenario.

\acknowledgments
We thank Omar Benhar, Pier Giuseppe Catinari,  Roberto Contino, Gabriele Franciolini, David Elazzar Kaplan, and Alessandro Strumia for useful conversations.
P.P. acknowledge financial support provided under the European
Union's H2020 ERC, Starting Grant agreement no.~DarkGRA--757480 and under the MIUR PRIN programme, and support from the Amaldi Research Center funded by the MIUR program ``Dipartimento di Eccellenza" (CUP:~B81I18001170001). The research of A.U. was supported in part by the MIUR under contract 2017\,FMJFMW (``{New Avenues in Strong Dynamics},'' PRIN\,2017).
This work was supported by the EU Horizon 2020 Research and Innovation Programme under the Marie Sklodowska-Curie Grant Agreement No. 101007855.

\appendix

\section{Embedding with the SM}\label{SM_derivation}

A simple way to derive Eq.~\eqref{theory_fund} is starting from the standard electroweak theory minimally coupled to Einstein gravity,
\begin{align}\label{starting_appendixA}
    S = \int \d^4 x\, \sqrt{-g}
    \Big[ \frac{R}{16\pi G} - \La_{\rm fields}\Big],
\end{align}
where
\begin{align}\label{La_fields}
    &\La_{\rm fields} =  - (\partial_\mu H)^\dagger (\partial^\mu H) - L^\dagger \slashed{D} L - R^\dagger \slashed{D}R \nonumber\\
    &- \frac{\lambda}{4}\Big(H^\dagger H - \frac{v^2}{2}\Big)^2 - f(L^\dagger H R + R^\dagger H^\dagger L),
\end{align}
and $H$ is the Higgs field, doublet under $SU(2)$, whereas
\begin{equation}
	L = \begin{pmatrix} v_L \\
		\psi_L
	\end{pmatrix},\quad R = \psi_R,
\end{equation}
are a $SU(2)$ doublet of left-handed fermions and a $SU(2)$ singlet of right-handed fermion, respectively. 

By exploiting the $SU(2)\times U(1)$ gauge symmetry to remove the spurious degrees of freedom, we write
\begin{equation}\label{key4}
	H = \begin{pmatrix}
		0 \\
		\frac{h}{\sqrt{2}}
	\end{pmatrix}.
\end{equation}
Substituting in Eq.~\eqref{starting_appendixA}, we recover Eq.~\eqref{theory_fund} where 
\begin{equation}
    \psi = \begin{pmatrix}
        \psi_L \\
        \psi_R
    \end{pmatrix}\,.
\end{equation}

\newpage 
\bibliography{refs}

\end{document}